\documentclass{hthp}
\usepackage[pdftex]{graphics}
\usepackage{supertabular}

\normalfont\upshape
\begin{document}

\title{The thermodynamics and transport properties of transition metals in critical point}
\author{Alexander L. Khomkin \and Aleksey S. Shumikhin \email{shum$_{-}$ac@mail.ru} }

\institute{Joint Institute for High Temperatures of RAS,
Izhorskaya Street, 13, bldg. 2, Moscow 125412, Russia}

\maketitle

\begin{abstract}
A new method for calculating the critical point parameters
(density, temperature, pressure and electrical conductivity) and
binodal of vapor-liquid (dielectric-metal) phase transition is
proposed. It is based on the assumption that cohesion, which
determines the main properties of solid state, also determines the
properties in the vicinity of the critical point. Comparison with
experimental and theoretical data available for transition metals
is made.
\end{abstract}


\section{Introduction}

Knowledge of the critical point parameters for a liquid-vapor
phase transition in metals is important both from the theoretical
standpoint and applied perspectives. It is particularly desirable
to distinguish vapors of transition metals, which are most widely
used for development of advanced structural materials and alloys.
The vapor-liquid phase transition in neutral gases (inert,
molecular, etc.) is well studied both theoretically and
experimentally \cite{Hirschfelder}. For the metal vapors, the situation is
different. The critical point parameters (density, temperature,
pressure, and electrical conductivity) and binodal are measured
only for alkali metals \cite{Hensel} and mercury \cite{Hensel-1995,Kikoin}. For most metals in
static experiments, the critical region is unattainable due to
high critical temperature ($\geq 10000$~K). Most of the
experiments are of a pulse nature. The data on the rapid pulse
heating of wires made of transition metals (Ti, V, Co, Fe, Cu, Mo,
Nb, Pd, W, Ir, Pt) in water and in inert gases under pressure
obtained by different authors are presented in \cite{Gathers,Pottlacher2010}. In these
studies, measurements of enthalpy, density, temperature and
electrical resistance were conducted within the temperature range
from the melting temperature $T_m$ to temperatures of $5000\div
7000$~K. In a number of works performed by the same method \cite{Hess,Beutl,Seydel,Boboridis,Martynyuk},
the authors were able to experimentally estimate the parameters of
critical points for several metals (V, Co, Fe, Mo, Au, Pt). For
refractory metals with high temperature and pressure at the
critical point (for example, Co, Fe, V), only the temperature and
pressure were measured \cite{Pottlacher}.

Some dynamic experiments on shock compression of porous elements
and their subsequent adiabatic expansion performed recently. The
measurements were made for nickel \cite{Nikolaev}, molybdenum \cite{Ternovoi}. Only
temperature and pressure were measured in these experimental
studies. It is not possible to determine the density at the
critical point and its vicinity (binodal) in this manner. It is
quite natural, that the measurement of conductivity in the
critical points and the near-critical branches of binodal are
completely absent.

Despite of the large number of theoretical methods estimating the
critical point parameters of metal vapors, a generally accepted,
unified approach (equations of state) is still not proposed.
Various theoretical estimates provide a substantial scatter of
data, especially for the temperature and pressure at the critical
point. Theoretical methods may be conditionally grouped as
follows: the use of empirical equations of state \cite{Young,Bushman,Likalter}, the
extrapolation of experimental data \cite{Martynyuk1998,Martynyuk1999,Blairs} in the near-critical
region and the use of general scaling and similarity laws,
established for neutral gases and liquids \cite{Grosse,Fortov,Vorobiev}.

The primary instrument of the majority of the above-mentioned
theoretical approaches is the extrapolation of particular values,
known in the vicinity of the melting point, to the critical
region. It might be either thermal values (pressure $P$, density
$\rho$, temperature $T$) -- thermal approaches or caloric values
(heat of evaporation -- $H_{evap}$, internal energy -- $E_{int}$) --
caloric approaches. For example, "thermal" and "caloric" models
for cesium and rubidium give very similar values of the critical
temperature and density. However, for transition metals, the
results provided by these models are very different from each
other. For example, according to various estimates for tungsten,
the range of the critical temperatures: $12000 < T_{cr} <
23000$~K.

The question of the electrical conductivity of metal vapors at the
critical point and its vicinity wasn't even considered in these
works. It seems to us that the electrical conductivity at the
critical point is an important characteristic of the vapor-liquid
phase transition in metal vapors, distinguishing it from phase
transition in neutral gases.

Calculations of thermodynamic and transport properties of metals
are usually performed independently from each other. Calculations
of the electrical conductivity of solid and liquid metals are
generally based on using of the Ziman formula \cite{Ziman}. For these
calculations, one needs to know the structure factor, scattering
cross sections on the ion cores, and the concentration of
conduction electrons (see., e.g., \cite{Apfelbaum,Redmer}). When approaching the
critical point, it is necessary to take into account the complex
effect of reducing the conduction electron concentration down to
zero. This effect, associated with the processes of localization
of conduction electrons on the ion cores, is hard for the
theoretical description. Theoretical calculation of the electrical
conductivity in this region is only possible via numerical methods
\cite{Levashov}. However, it should be noted that existing software packages
(e.g., WASP) do not allow to perform simultaneous calculations of
the two-phase region boundaries and electrical conductivity.

We have published a series of papers \cite{Khomkin2014,Khomkin2015,Khomkin2016}, devoted to the
calculation of the critical point parameters of the vapor-liquid
(dielectric-metal) phase transition. Method of calculation is
based on a physical model with the hypothesis of the decisive role
of cohesion (the collective quantum binding energy) during the
formation of a liquid metallic phase. The critical point
parameters for most metals, semiconductors, and inert gases \cite{Khomkin2016}
were calculated using the constructed equation of state. The
universality of the cohesion was shown: this energy is quantum for
metals and classic for inert gases.

In this paper, thermodynamic parameters and electrical
conductivity of vapors of transition metals at the critical point
are calculated on the basis of the proposed equations of state
(EOS) and the Regel-Ioffe formula for the minimum metallic
conductivity. Calculating the concentration of conduction
electrons (Bloch electrons) at the critical point is performed by
two methods: using scaling relations \cite{Smith,Rose}, and using the
calculation of the jellium density (EAM -- Embedded Atom Method)
with use of Hartree-Fock-Slater wave functions of an isolated atom
\cite{Clementi}. The comparison made with the estimates for the critical
point parameters of other authors.

\section{The main relations}


The accurate calculation of cohesion is only possible for metals
with one valence s-electron. For metals with many-electron outer
shell, computation of the binding energy is rather time-consuming
task. We used a universal ratio for binding energy
$E_{UBER}(\Delta E, a^{\ast})$ (UBER -- Universal Bind Energy
Relation), proposed in \cite{Rose}. The ratio summarizes the analytical
data of numerous numerical calculations and describes quite well
the different types of bonding energy in metals in dependence on
the density of atoms $n_a = N_a/V$ and some physical quantities,
characteristic only for the substances in question.

\begin{equation} 
 \label{eq1}
 E_\mathrm{coh}(r_{WS}) = \Delta E E^{\ast}(a^{\ast});
\end{equation}

\begin{equation} 
 \label{eq2}
 E^{\ast}(a^{\ast}) = -(1+a^{\ast})\exp(-a^{\ast});
\end{equation}

\begin{equation} 
 \label{eq3}
 a^{\ast} = \frac{r_{WS} - r_{WSE}}{l}.
\end{equation}
Here $E^{\ast}(a^{\ast})$ is the UBER; $a^{\ast}$ is the scaled
interatomic separation; $r_{WS}$ is the radius of the Wigner-Seitz
cell; $r_{WSE}$ is the equilibrium radius, corresponding to the
solid state density; $\Delta E$ is the equilibrium binding energy;
$l$ is the scaling length, binding with the isothermal bulk
modulus $B$ by relation: $l = (\Delta E/(12\pi B r_{WSE}))^{1/2}$.
The unitless scaling parameters in Bohr radius are: $l_0 = l/a_0$,
$y_0 = r_{WS0}/a_0$, $y = r_{WS}/a_0$. As a result, the binding
energy depends on the current density ($y$) and three parameters
($\Delta E$, $y_0$, $l_0$) -- $E_{UBER}(\Delta E, y_0,l_0,y)$.
These data are tabulated in \cite{Rose}.

The Helmholtz free energy for $N_a$ atoms in volume $V$ at
temperature $T$, proposed in \cite{Khomkin2014}, has the form:

\begin{equation} 
\label{eq4}
\beta F =
-N_{a}ln\frac{eVg_a}{N_{a}\lambda_{a}^{3}}+N_{a}\frac{4\eta -
3\eta^{2}}{(1-\eta)^{2}}+\frac{1}{2}\beta N_{a}E_\mathrm{coh}(y) ,
\end{equation}
where $N_a$, $\lambda_{a}$ and $g_{a}$ are the concentration of
atoms, the thermal de Broglie wavelength and the statistical
weight of an atom, respectively; $\beta = 1/(k_{B}T)$ is the
inversed temperature, $k_B$ is the Boltzmann constant, $T$ is the
temperature in Kelvins. $\eta = 4/3 \pi n_{a}R_{a}^{3}$ is the
packing parameter, where $R_{a}=r_{WSE}-l$ is the radius of an
atomic hard sphere; $n_{a}$ is the number density of atoms.

In \cite{Smith,Rose}, one may find the figures, which compares numerous
data of numerical calculations of cohesion and calculation using
UBER. They all demonstrate a high accuracy (within $1-2 \%$) of
the scaling function.

\section{The density of conduction electrons}

The density of conduction electrons of metals in solid or liquid
state is defined by the zone structure and the total concentration
of such electrons, which in turn is determined by the product of
the effective charge of the ion core and the concentration of
nuclei. For most metals under normal conditions, these data are
practically tabulated and do not require computational efforts,
that is used in calculating the conductivity of solid and liquid
metals. The situation changes completely at depression of metal or
at approach to the critical point. Difficult processes of return
of Bloch electrons into nuclear orbits with a formation, finally,
of neutral atoms into which liquid metal at an exit from
near-critical SCF of area turns begin. The effective charge of the
ion core $\alpha$ (it also describes the degree of the "cold
ionization") continuously strives to zero, which means the
disappearance of the Bloch conduction electrons. It should be
noted that the strict calculation of the degree of the "cold
ionization" for such a phase transition is a complicated task that
requires the involvement of methods of density functional and
molecular dynamics, implemented, for example, in the WASP package.
But even with use of these modern numerical methods such
calculations aren't always possible, particularly in the vicinity
of a critical point where the two-phase area appears and the
system becomes strongly non-uniform. The technique of analytical
estimates offered in this work can be useful, especially in a
situation when both experimental and systematic settlement data
are absent. Using the theory of Bardeen \cite{Bardeen} for the perfect
lattice of atoms with one valence s-electron (the hydrogen and
vapors of alkali metals), the dependence of the effective charge
$\alpha$ on the density can be calculated analytically, which
allowed in \cite{Khomkin2014} to make a preliminary estimates in the critical
point. The value of $\alpha$ is about $1/3$. Upon further
depression the value $\alpha$ goes to zero, but not abrupt, rather
gradually.

\subsection{Determining of the electron density using Scaling}

The development of the EAM \cite{Daw} led to numerous calculations of
the binding energy of an arbitrary atom immersed in jellium of
arbitrary nature, formed by either the same or other atoms. In
\cite{Smith}, these data were processed, generalized and presented in the
form of universal scaling dependencies of the atom's binding
energy both on the density of the nuclei and the density of
jellium. Based on the equality of these dependencies, a formula
was proposed which links a unitless value of the jellium density
with the parameter $a^{\ast}$:

\begin{equation} 
 \label{eq5}
 \frac{n_{e}}{n_m} = (e^{a^{\ast}})^{1/\gamma} = \alpha_{Sc}.
\end{equation}

Here $n_e$, $n_m$ are the current electron concentration and the
electron concentration in the metal at normal density,
respectively; the exponent $\gamma = \lambda_{TF}/l$ where the
value $l$ is the same as in~(\ref{eq3}), and $\lambda_{TF}$ is the
Thomas-Fermi screening length \cite{Rose1983}:

\begin{equation}
 \label{eq6} 
 \lambda_{TF} = \frac{1}{3}\left(\frac{243\pi}{64}\right)^{1/6}n_{m}^{-1/6}.
\end{equation}

The concentration of electrons in a metal at normal density $n_m$
is the tabular value. As a rule, it may be associated with a
valence $Z$ and density of the metal nuclei under normal
conditions $n_0$ by the ratio $n_m = Z n_0$.

A way of calculation of the electron density at the depression of
metal nuclei based on the ratio (\ref{eq5}) is named below as "scaling".

\subsection{The calculation of the electron density using Hartree-Fock-Slater wave functions}

In \cite{Clementi}, the data are presented of the wave functions of an
isolated atom, calculated numerically by the Hartree-Fock method
and presented in the form of expansions of Slater-type orbitals.
The data cover all elements up to the atomic number $Z = 54$.

The wave function $\Psi(r)$ of an arbitrary i-th atomic electron
in a particular quantum state, appears in the form of expansion of
the Slater-type orbitals $\chi_{\lambda p}(r,\theta,\varphi)$:

\begin{equation}
 \label{eq7}
 \Psi({\bf r}) = \sum_{\lambda, p} C_{\lambda, p} \chi_{\lambda,
 p}(r,\theta,\varphi).
\end{equation}

Knowing the wave function of the i-th electron of the isolated
atom, we can calculate the proportion of the electron density
involved in the formation of jellium in an cell approximation. In
the EAM method, the fraction $\alpha_{HF}$ is determined by
integrating the $(\Psi(r))^2$ outside the Wigner-Seitz cell and
the contribution of the permanent background within the cell
$(\Psi(y))^2$:

\begin{equation} 
 \label{eq8}
 \alpha_{HF}^{i} = \int_{y}^{\infty} (\Psi(r))^{2}r^2 dr +
 \frac{y^3}{3}(\Psi(y))^{2} ,
\end{equation}
where $y$ is the current radius of the Wigner-Seitz cell in atomic
units. Slater-type orbitals $\chi_{\lambda p}(r,\theta,\varphi)$
are written as a product of normalized to unit of standard radial
and spherical functions. Constants for calculation Slater-type
orbitals (\ref{eq7}) are presented in \cite{Clementi} in the form of tables for all
electronic states. Formally, we can calculate the value
$\alpha_{HF}$ for all electrons of the atom. Their sum give the
estimate of the sought-for degree of the "cold ionization". In our
computations, we used the data from \cite{Clementi}, but only for valence
electrons, because the electron contribution of the ion core in
our conditions is small and does not affect the ultimate value
$\alpha_{HF} = \sum_{i} \alpha_{HF}^i$. Moreover, keep in mind,
that in the vicinity of the critical point even the valence
electrons participate only partially in the formation of the
jellium. When approaching the normal density of the metal, all the
valence electrons are involved in formation of jellium and
$\alpha_{HF}$ strives to the total valence. The electrons of the
ion core are involved in formation of the jellium only upon
further compression.

Concentration of conductivity electrons in this calculation option
is defined by a ratio:

\begin{equation}
 \label{eq9}
 n_e = \alpha_{HF} n_0 .
\end{equation}

We will call this variant the "Hartree-Fock".

\subsection{The calculation of electrical conductivity}

The mean path length of the conduction electrons ($l_p$) is
inversely proportional to the product of the scattering cross
sections and the structural factor. In the calculation of the
electrical conductivity at the critical point, we use the fact
that $l_p$ decreases with the metal depression due to the growth
of the structure factor associated with the loss of long-range
order. In an assumption that the mean path length cannot be less
than the average interparticle distance $l_p \geq 2r_{WS}$, a
simple formula was proposed for the minimum conductivity of metals
by Regel-Ioffe. This formula doesn't contain a product of the
structural factor and the scattering cross sections, and uses only
the minimum path length $l_p = 2 r_{WS}$. It seems quite
reasonable the use of such approximation at the critical point:

\begin{equation}
 \label{eq10}
 \sigma = n_{e} \frac{q_{e}^{2}}{m_e} \tau,
\end{equation}
where $q_e$ and $m_e$ are the charge and the mass of an electron,
respectively; $\tau$ is the mean free time. The value $\tau$ is
defined as transit flight time of internuclear distance, which is
equal to the doubled radius of the Wigner-Seitz cell $2r_{WS}$
($2y$ in atomic units), with Fermi velocity $v_F = p_{F}/m_{e}$:

\begin{equation} 
 \label{eq11}
 \frac{\tau}{m_e} = \frac{2r_{WS}}{p_F} ,
\end{equation}
where $p_F = (3\pi^2 n_e)^{1/3}\hbar$ is the Fermi momentum. As a
result, we obtain the following expression for the electrical
conductivity:

\begin{equation} 
 \label{eq12}
 \sigma = n_{e}^{2/3}
 \frac{q_{e}^{2}}{9\cdot 10^{11}}\frac{2y}{(3\pi^2)^{1/3}\hbar}.
\end{equation}

The minimum metallic conductivity in the vicinity of the critical
point is determined by the concentration of conduction electrons
$n_e$, related to the nuclei density both by ratios (\ref{eq5},\ref{eq9}) and a
direct dependency on the nuclei density via $y$ -- radius of the
cell in atomic units. The temperature dependence is absent.
Dimensions of all quantities in (\ref{eq12}) -- CGSE, and the conductivity
is in $1/(\Omega cm)$.

\section{Results and Discussion}

Using the expression (\ref{eq4}) for the Helmholtz free energy, we
obtained the equation of state and calculated isotherms for
various substances. Here, we consider isotherms similarly to \cite{Khomkin2014,Khomkin2015}, by plotting a curve representing the pressure-density
dependence. As the temperature decreases, the isotherms
demonstrate the appearance of the van der Waals loop, which
clearly indicates the presence of the first-order vapor-liquid
phase transition. Analyzing the isotherms, we can estimate values
of all three critical parameters: temperature, density and
pressure.

Table~\ref{tab11} presents the obtained parameters of the critical
points. This table also presents estimates by other authors: using
the method of corresponding states \cite{Fortov}; using various
modifications of the van der Waals equation of state \cite{Gathers,Young,Martynyuk1998,Martynyuk1999}; scaling relations and models of virtual atoms by Likalter
\cite{Likalter2002,Likalter1997}; an experimental
estimation using a pulse heating of the wires \cite{Hess,Beutl,Seydel,Boboridis,Martynyuk,Pottlacher}; an
experimental estimation using dynamic compression of porous
materials \cite{Nikolaev,Ternovoi}. The scatters of estimates for density and,
especially, for temperature and pressure are very large. Part of
the methods mentioned above use different experimental data on the
melting curve, from the melting point to the boiling point.
Thermal and caloric approaches provide the significantly different
parameters of critical points. Note, that we do not use
experimental data on the melting curve in our model. For
calculations using the equation of state (\ref{eq1}~--~\ref{eq4}), we need values
of the heat of evaporation, the normal density and the isothermal
bulk modulus for the metal in a solid state. These values with
known high accuracy presented in \cite{Rose}.

The table~\ref{tab11} shows that for the static experiments with
most of the metals the temperatures at the critical point are too
high. The table shows available data for the critical temperature
and pressure, experimentally measured in the dynamic experiments
by adiabatic expansion of the porous metal after the shock
compression \cite{Nikolaev,Ternovoi}. Unfortunately, measurements of the critical
density in this way are not yet possible.


\tablecaption{The calculated critical
parameters}
\tablehead{\hline} 
\tabletail{\hline} 
\label{tab11}
\begin{supertabular}{|c|c|c|c|c|}
\hline

Metal & $\rho_{cr}$, g/cm$^{3}$ & $T_{cr}$, K & $P_{cr}$, MPa & Ref \\ \hline

Ti & 1.31 & 11790 & 763 & \cite{Fortov} \\ 
--- & 0.67 & 9040 & 156 & \cite{Martynyuk1999} \\ 
--- & 1.05 & 10700 & 1150 & this work \\ \hline

V & 1.56 & 11325 & 1031 & \cite{Young} \\ 
 --- & 1.86 & 12500 & 1078 & \cite{Fortov} \\  
 --- & 1.63 & 6396 & 920 & \cite{Gathers} \\  
 --- & 1.55 & 8550 & 648 & \cite{Martynyuk} \\  
 --- & 0.91 & 9980 & 223 & \cite{Martynyuk1999}  \\ 
 --- & 1.4 & 11600 & 1620 & this work \\ \hline

Cr &   & 10500 & 935 & \cite{Likalter} \\ 
 --- & 2.22 & 9620 & 968 & \cite{Fortov} \\ 
 --- & 2.0 & 8000 & 1660 & this work \\ \hline

Fe & 2.03 & 9600 & 825 & \cite{Fortov} \\
--- & 2.04 & 9340 & 1035.4 & \cite{Young} \\
--- & 1.63 & 7650 & 153.4 & \cite{Filippov} \\
--- & --- & 9250 $\pm$ 700 & 875 $\pm$ 50 & \cite{Beutl} \\ 
--- & 1.4 & 7928 & 285.8 & \cite{Vorobiev} \\
--- & 2.31 & 10637/5433 & 1253/657 & \cite{Apfelbaum} \\ 
--- & 1.296 & 8310 & 272 & \cite{Martynyuk1998} \\ 
--- & 1.98 & 8950 & 1610 & this work \\
\hline

Co & 2.2 & 10460 & 923 & \cite{Fortov} \\ 
--- & --- & 10384 $\pm$ 700 & 1106 $\pm$ 60 & \cite{Hess} \\ 
--- & 2.2 & 8950 & 1820 & this work \\ \hline

Ni & 2.3 & 9600 & 1100 & \cite{Young} \\ 
--- & 2.19 & 10330 & 912 & \cite{Fortov} \\ 
--- &   & 11500 & 1500 & \cite{Likalter} \\ 
--- & 1.37 & 8554 & 269.4 & \cite{Martynyuk1998} \\ 
--- & --- & 9100 $\pm$ 150 & 900 $\pm$ 100 & \cite{Nikolaev} \\ 
--- & 2.2 & 9300 & 1820 & this work \\ \hline

Cu & 2.33 & 7600 & 830 & \cite{Young} \\ 
--- & 2.39 & 8390 & 746 & \cite{Fortov} \\ 
--- & 1.94 & 8440 & 651  & \cite{Likalter2002} \\ 
--- & 1.95 & 7093 & 45 & \cite{Vorobiev2009} \\ 
--- & 1.58 & 7580 & 800 & \cite{Vorobiev} \\ 
--- & 2.19 & 5890 & 169 & \cite{Martynyuk} \\ 
--- & 2.3 & 7250 & 1350 & this work \\ \hline

Zn & 2.29 & 3190 & 263 & \cite{Fortov} \\ 
--- & 2.0 & 3170 & 290 & \cite{Young} \\ 
--- & 2.62 $\pm$ 0.52 & 3600 $\pm$ 600 & 350 $\pm$ 30 & \cite{Pottlacher} \\ 
--- &  & 3620 & 246 & \cite{Martynyuk} \\ 
--- & 1.733 & 3485 & 199 & \cite{Martynyuk1998} \\ 
--- & 2.25 & 2120 & 540 & this work \\ \hline

Y & 0.566 & 7510 & 161 & \cite{Likalter2002} \\ 
--- & 1.1 & 9500 & 600 & \cite{Likalter1997} \\ 
 --- & 1.3 & 10800 & 374 & \cite{Fortov} \\ 
 --- & 1.0 & 10300 & 500 & this work \\  \hline

Zr & 1.79 & 16250 & 752 & \cite{Fortov} \\
--- & 1.4 $\pm$ 0.3 & 14500 $\pm$ 1500 & 410
& \cite{Onufriev} \\ 
--- & 2.24 & 9660 & 667.4 & \cite{Vorobiev} \\
--- & 0.84 & 10720 & 102 & \cite{Martynyuk1999} \\ 
--- & 1.4 & 14400 & 1070 & this work \\
\hline

Nb & 2.59 & 19040 & 1252 & \cite{Fortov} \\ 
--- & 2.02 & 9989 & 963 & \cite{Gathers} \\ 
--- & 1.04 & 12320 & 138 & \cite{Martynyuk1999} \\ 
--- & 2.02 & 11200 & 607 & \cite{Martynyuk} \\ 
--- & 2.0 & 16200 & 1760 & this work \\ \hline

Mo & 2.62 & 14588 & 1184.4 & \cite{Young} \\
--- & 3.18 & 16140 & 1263 & \cite{Fortov} \\
--- & 2.3 & 8002 & 970 & \cite{Gathers} \\
--- &  & 12500 $\pm$ 1000 & 1000 $\pm$ 100 &
\cite{Ternovoi} \\ 
--- & 2.63 & 11150 & 546 & \cite{Seydel} \\
--- & 2.47 & 10780 & 692 &
\cite{Martynyuk} \\ 
--- & 1.37 & 11330 & 175 & \cite{Martynyuk1999}
\\ 
--- & 2.8 & 12870 & 2240 & this work \\
\hline

Ru & 3.79 & 15500 & 1374 & \cite{Fortov} \\ 
 --- & 3.48 & 12180 & 2580 & this work \\ 

Pd & 3.06 & 8301 & 708.5 & \cite{Young} \\ 
 --- & 3.2 & 10760 & 764 & \cite{Fortov} \\ 
 --- & 3.5 & 6850 & 1490 & this work \\ \hline

 Ag & 2.7 & 6410 & 480 & \cite{Young} \\ 
 --- & 2.93 & 7010 & 450 & \cite{Fortov} \\ 
 --- & 3.0 & 5500 & 900 & this work \\ \hline

Cd & 2.33 & 2619 & 161.5 & \cite{Young} \\ 
 --- & 2.74 & 2790 & 160 & \cite{Fortov} \\ 
 --- & 3.0 & 1600 & 360 & this work \\ \hline

Re & 5.4 & 17293 & 1488 & \cite{Young} \\ 
 --- & 6.32 & 19600 & 1570 & \cite{Fortov} \\  
 --- & 4.4 & 11500 & 1400 & \cite{Likalter1997} \\  
 --- & 2.7 & 13070 & 195 & \cite{Martynyuk1999}  \\ 
 --- & 6.1 & 14600 & 2940 & this work \\ \hline

Ir & 5.64 & 10340 & 950 & \cite{Gathers} \\ 
 --- & 6.77 & 15380 & 1278 & \cite{Fortov} \\  
 --- & 2.98 & 12120 & 208 & \cite{Martynyuk1999} \\ 
 --- & 6.64 & 11900 & 2790 & this work \\ \hline

 Pt & 5.5 & 12526 & 1050.5 & \cite{Young} \\ 
 --- & 5.02 & 14330 & 870 & \cite{Fortov} \\ 
 --- & 4.72 & 9286 & 949 & \cite{Gathers} \\ 
 --- & 5.08 & 8970 & 388 & \cite{Martynyuk} \\ 
 --- & 2.85 & 10450 & 172 & \cite{Martynyuk1999} \\ 
 --- & 6.2 & 10150 & 2200 & this work \\ \hline

Au & 5.0 & 8267 & 626.5 & \cite{Young} \\ 
 --- & 5.68 & 8970 & 610 & \cite{Fortov} \\ 
 --- & 4.35 & 8100 & 462 & \cite{Morris} \\  
 --- & 7.7 $\pm$ 1.7 & 7400 $\pm$ 1100 & 530 $\pm$ 20 & \cite{Boboridis} \\ 
 --- & 6.1 & 6250 & 1290 & this work \\ \hline

\end{supertabular}



\begin{figure}[tbp]
\begin{center}

\scalebox{0.5}{\includegraphics{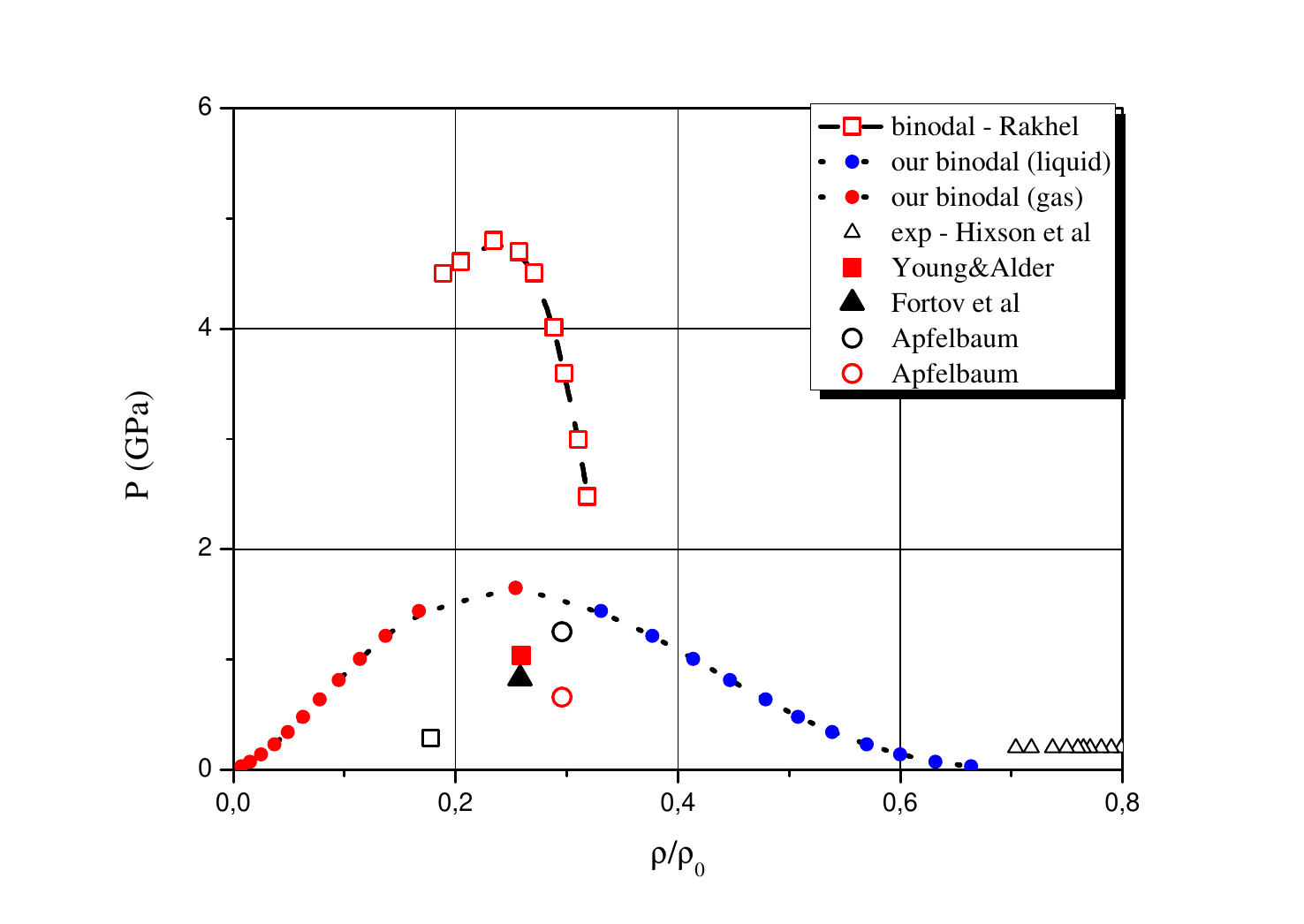}} a)

\scalebox{0.5}{\includegraphics{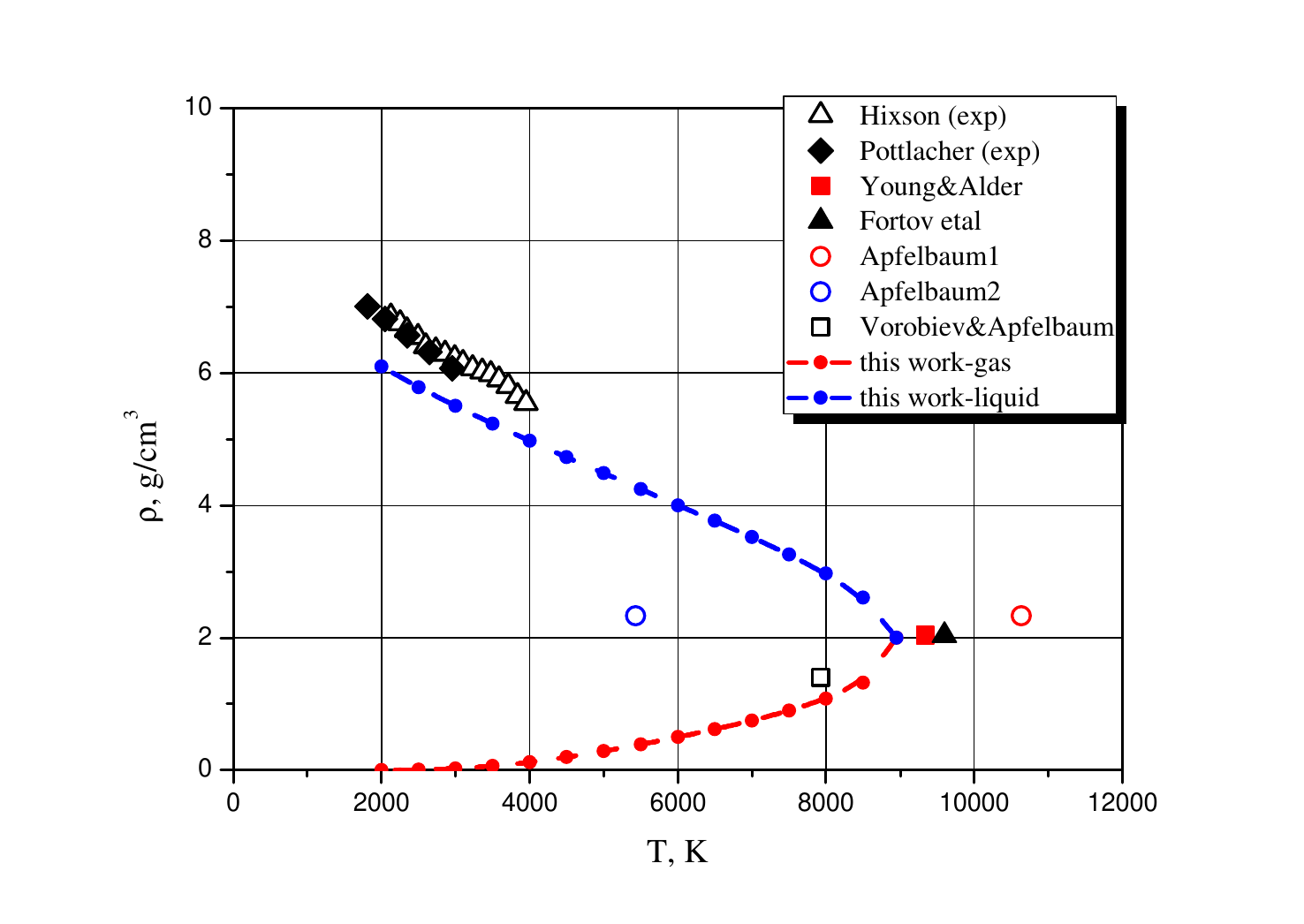}} b)
\end{center}
\caption{Binodal of iron in coordinates $P$ vs $\rho/\rho_0$ (1a) and $\rho$ vs $T$ (1b), respectively. Dashed line with points --
this work. Experimental data: open triangles -- Hixson \cite{Hixson}, dashed line with open squares -- Rakhel
et al \cite{Rakhel}, diamonds -- Beutl et al \cite{Beutl}. Theoretical
estimations of critical point: filled square -- Young and Alder
\cite{Young}, filled triangle -- Fortov et al [22], open square --
Vorob'ev and Apfelbaum \cite{Vorobiev}; open circles -- Apfelbaum \cite{Apfelbaum2011}.}
\label{Fe1a}
\end{figure}

\begin{figure}[tbp]
\begin{center}
\scalebox{0.5}{\includegraphics{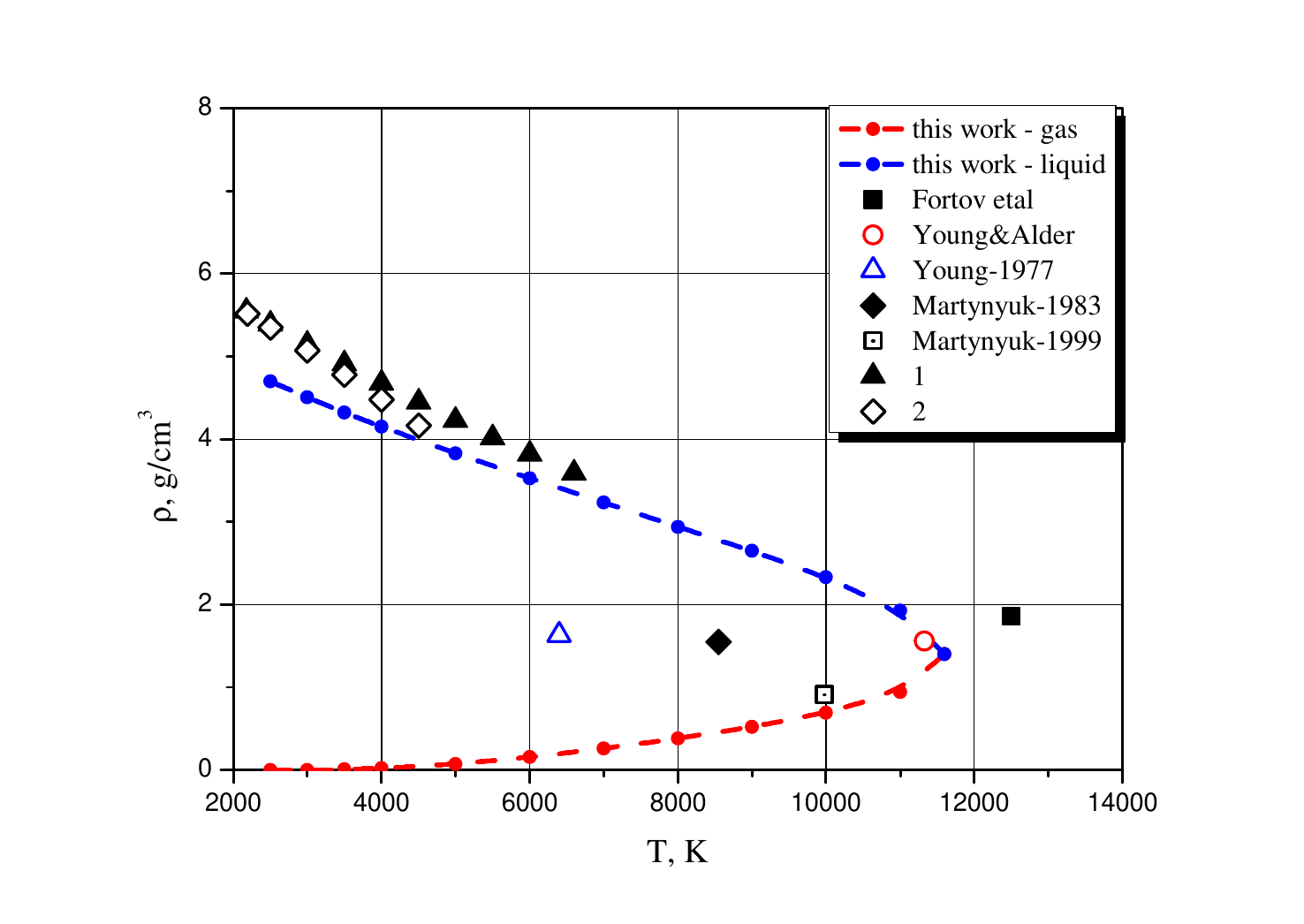}}
\end{center}
\caption{Binodal of vanadium in coordinates $\rho$ vs $T$. Dashed
line with points -- this work. Experimental data: 1 -- data from
fast system at Kiel \cite{Gathers}; 2 -- data at $0.3$~GPa from fast system
at Livermore \cite{Gathers}. Theoretical estimations of critical point: open
circle -- Young and Alder \cite{Young}; filled square -- Fortov etal \cite{Fortov};
open triangle -- estimate of CP using van der Waals model with
soft spheres, Young, 1977 \cite{Gathers}; filled diamond -- estimate from
experimental data by Martynyuk \cite{Martynyuk}.} \label{Vanadium}
\end{figure}

\begin{figure}[tbp]
\begin{center}
\scalebox{0.5}{\includegraphics{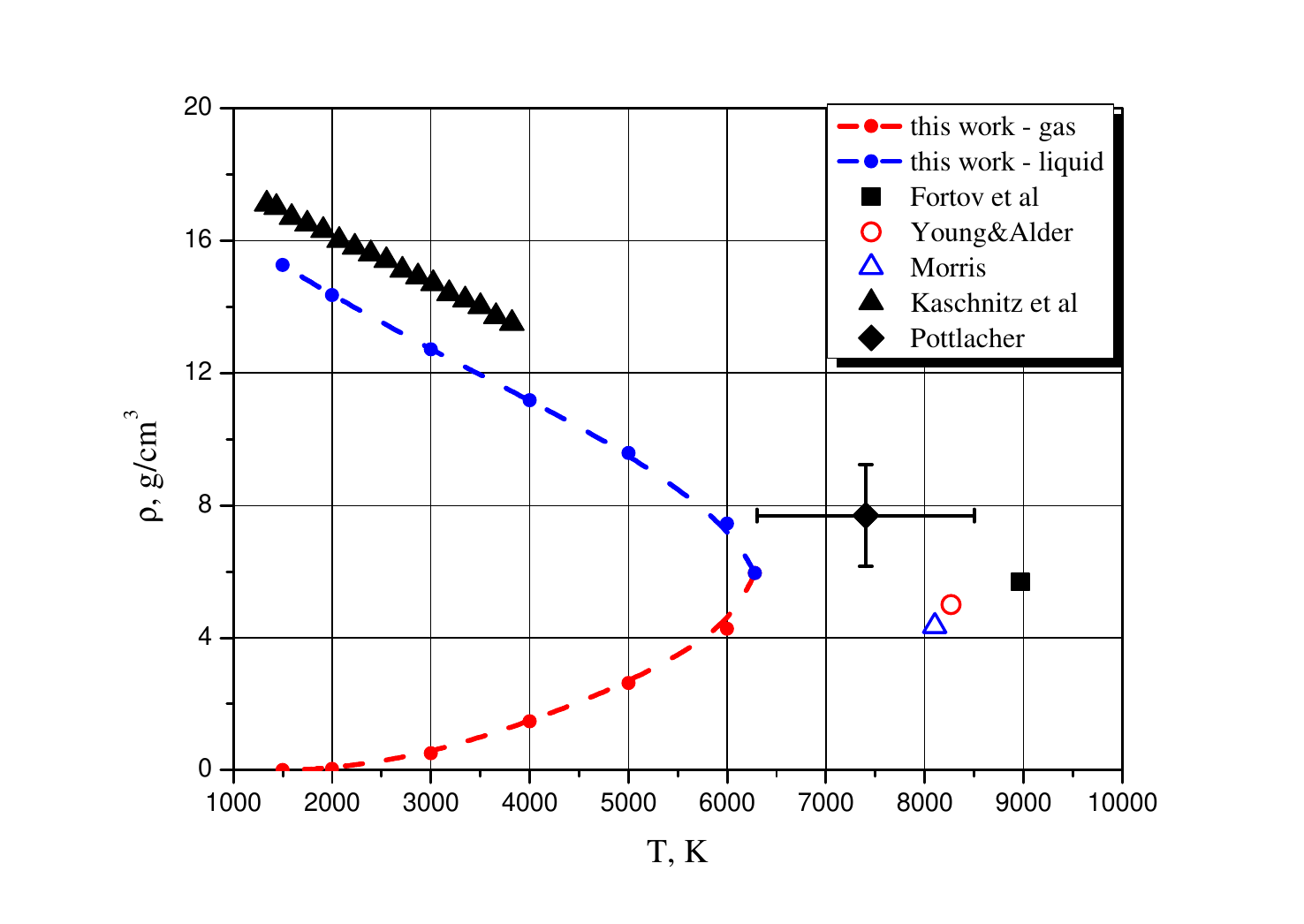}}
\end{center}
\caption{Binodal of gold in coordinates $\rho$ vs $T$. Dashed line
with points -- this work. Diamond with error bar -- experimental
estimate of CP by Pottlacher et al \cite{Boboridis}. Filled triangles --
experimental data of Kaschnitz et al \cite{Kaschnitz}. Open circle -- Young
and Alder \cite{Young}; filled square -- Fortov et al \cite{Fortov}; open triangle --
Morris \cite{Morris}.} \label{Au}
\end{figure}

There is a large number of theoretical and experimental estimates
of critical parameters for transition metals, especially for Co,
V, Fe, and Mo. Despite that, no unified model widely accepted for
calculation of the critical point of the vapor-liquid phase
transition and, especially, for calculation of binodal of this
transition, is available yet. Existing experimental data allow us
to estimate the temperature and pressure at the critical point,
but to assess the density is still quite difficult. Therefore, to
restore the coexistence curve of phases is experimentally not
possible. Our model allows calculating the binodal of the
vapor-liquid phase transition for any metal.
Figures~\ref{Fe1a}--\ref{Au} show the binodal for iron, vanadium,
and gold, respectively.

Figure~\ref{Fe1a} presents binodal of iron in the coordinates $P$ vs
$\rho/\rho_0$ (1a) and $\rho$ vs $T$ (1b),
respectively. Triangles correspond to the experimental data on
$0,2$~GPa isobar from Hixson \cite{Hixson}, diamonds correspond to the experimental data obtained by the pulse heating of wires from
work Beutl et al \cite{Beutl}. Squares correspond to the
binodal, obtained experimentally in work of Rakhel with co-authors
\cite{Rakhel}. Theoretical estimations of critical point parameters
obtained by various authors are also shown: the filled square
correspond to the calculation using the van der Waals equation of
state from Young and Alder \cite{Young}; triangle -- calculation using the
method of corresponding states by Fortov et al \cite{Fortov}; open square --
calculation using the similarity law from the work of Vorob'ev and
Apfelbaum \cite{Vorobiev}; open circles -- calculation using the Morse
potential from the work of Apfelbaum \cite{Apfelbaum2011}. The dashed curve with dots marks our
calculation. As can be seen from figure~\ref{Fe1a}, our calculations for
the liquid branch of the binodal are in good agreement with
available experimental data \cite{Beutl}.

Figure~\ref{Vanadium} presents binodal of vanadium in the
coordinates $\rho$--$T$. Experimental data from fast pulse heating
systems are shown: 1 -- data at Kiel \cite{Gathers}; 2 -- data at $0,3$~GPa
at Livermore \cite{Gathers}. Theoretical estimates of the critical point,
obtained by various authors: open circle -- calculation using the
van der Waals equation of state by Young and Alder \cite{Young}; filled
square -- calculation using the method of corresponding states
from Fortov et al \cite{Fortov}; filled triangle -- calculation using the
van der Waals equation of state with soft spheres by Young, 1977
\cite{Grosse}; filled diamond -- estimate from experimental data by
Martynyuk \cite{Martynyuk}.

Figure~\ref{Au} presents binodal of gold in the coordinates
$\rho$--$T$. The filled triangles correspond to the experimental
data on the wire explosion from Kaschnitz et al \cite{Kaschnitz}. Diamond with
error bars -- experimental estimation of the critical point from
Pottlacher et al \cite{Boboridis}. Theoretical estimates of the critical point,
obtained by various authors: open circle -- calculation using the
van der Waals equation of state by Young and Alder \cite{Young}; filled
square -- calculation using the method of corresponding states
from Fortov et al \cite{Fortov}; open triangle -- Morris \cite{Morris}.

As can be seen from the figures, the results of our calculations
in the framework of a rather simple physical model quite well
agree with known experimental data for liquid metals in the
temperature range from the melting temperature to $T \sim 5000$~K.
Various estimations of the critical density for iron give similar
values. However, the scatter of estimates for the critical
temperature and, especially, for the critical pressure of iron is
large. This situation is typical for most transition metals. Our
calculations clarify the available data.

As an example, Table~\ref{tab14} presents calculations of the
electrical conductivity at the critical point (critical electrical
conductivity) for Cu, Ag, Fe, V, and Zn, performed by the
Regel-Ioffe formula (\ref{eq12}), but with different values of the degree
of "cold ionization" $\alpha$. The index of the symbol of the
electrical conductivity $\sigma$ indicates the method of
calculation of the $\alpha$ value. Available estimation of the
electrical conductivity for iron \cite{Rakhel} at the critical point is
also presented in Table~\ref{tab14}.

\begin{table}[tbp]
\caption{Electrical conductivity at the critical point for various
metals }
\begin{center}
\begin{tabular}{|@{\hspace{1mm}}c|@{\hspace{1mm}}c|@{\hspace{1mm}}c|@{\hspace{1mm}}c|}
\hline
Metal & $\sigma_{Sc}$, 1/($\Omega$cm) & $\sigma_{HF}$,
1/($\Omega$cm) & $\sigma_{exp}$, 1/($\Omega$cm) \\ \hline
V & 2540
& 2440 &  \\ \hline
Zn & 3100 & 2040 &  \\ \hline
Cu & 2520 & 1660
& \\ \hline
Ag & 2180 & 1570 &  \\ \hline
Fe & 3440 & 2560 & 2500
$\pm$ 800 \\ \hline
\end{tabular}%
\end{center}

\label{tab14}
\end{table}

We have performed calculations of the parameters of the critical
point for almost all metals of the periodic table. Analysing the
obtained results, we noticed some interesting regularities. The
unitless parameter $a^{\ast}$ at the critical point is with good
precision ($\sim 10 \%$) close to $3$ for most metals. From this
fact, a simple relation follows for the unitless radius of the
Wigner-Seitz cell at the critical point $y_{cr}$, that allows
calculating the critical density:

\begin{equation}
 \label{eq13}
 y_{cr} = y_0 + 3l_{0}.
\end{equation}

From our model, it follows that the ratio of cohesion in the
minimum $\Delta E$ (heat of evaporation under normal conditions)
to the critical temperature is $\sim 5 \div 6$ , thus, follows the
Kopp-Lang rule \cite{Lang}:

\begin{equation} 
 \label{eq14}
 \Delta E/T_{cr} = const = 5\div 6.
\end{equation}

Relations (\ref{eq13}, \ref{eq14}), obtained "theoretically", are peculiar
similarity relations binding the critical density and temperature
with the evaporation heat, the normal density, and the isothermal
bulk modulus.

\section{Conclusions}

In this paper, we propose a model for calculating the parameters
of the critical point and electrical conductivity, as well as
binodal of the vapor-liquid phase transition for transition metals
(metals with an uncompleted outer electron shell). The model is
based on the hypothesis about the decisive role of collective
quantum binding energy -- cohesion for the description of the
interatomic interactions for the metal both in the condensed
state, and in the gas state near the critical point. To calculate
cohesion of multielectron atoms of the metals, we suggest using
the scaling relations known from the literature. The parameters of
the critical point are obtained, and the binodals are calculated
for transition metals. The critical electrical conductivity for
most metals is calculated for the first time.

Final confirmation of the accuracy of calculations within
"thermal" or "caloric" approaches requires carrying out further
experiments. Most likely, it will be experimenting with
installations on shock compression and the subsequent adiabatic
expansion of porous samples of metals.

\section{Acknowledgments}

This study was supported by the Russian Science Foundation under
the Project No 14-50-00124.

\bibliographystyle{hthp}
\bibliography{hthp2}

\begin{thebibliography}{10}
\providecommand{\url}[1]{\texttt{#1}}
\providecommand{\urlprefix}{URL }
\expandafter\ifx\csname urlstyle\endcsname\relax
  \providecommand{\doi}[1]{doi:\discretionary{}{}{}#1}\else
  \providecommand{\doi}{doi:\discretionary{}{}{}\begingroup
  \urlstyle{rm}\Url}\fi

\bibitem{Hirschfelder}
Hirschfelder J.O., Curtiss C.F., Byron~Bird R.
\newblock \emph{Molecular Theory of Gases and Liquids}.
\newblock New York: Wiley, 1954.

\bibitem{Hensel}
Hensel F., Marceca E., Pilgrim W.C.
\newblock \emph{J Phys-Condens Mat} \textbf{10} (1998) 11395.

\bibitem{Hensel-1995}
Hensel F.
\newblock \emph{Adv Phys} \textbf{44} (1995) 3.

\bibitem{Kikoin}
Kikoin I.K., Senchenkov A.P.
\newblock \emph{Fiz Met Metalloved+} \textbf{24} (1967) 843.

\bibitem{Gathers}
Gathers G.R.
\newblock \emph{Rep Prog Phys} \textbf{49} (1986) 341.

\bibitem{Pottlacher2010}
Pottlacher G.
\newblock \emph{High temperature thermophysical properties of 22 pure metals}.
\newblock Graz: Edition Keiper, 2010.

\bibitem{Hess}
Hess H., Kaschnitz E., Pottlacher G.
\newblock \emph{High Pressure Res} \textbf{12} (1994) 1323.

\bibitem{Beutl}
Beutl M., Pottlacher G., J{\"a}ger H.
\newblock \emph{Int J Thermophys} \textbf{15} (1994) 1323.

\bibitem{Seydel}
Seydel U., Fucke W.
\newblock \emph{J Phys F Met Phys} \textbf{8} (1978).

\bibitem{Boboridis}
Boboridis K., Pottlacher G., J{\"a}ger H.
\newblock \emph{Int J Thermophys} \textbf{20} (1999) 1289.

\bibitem{Martynyuk}
Martynyuk M.M.
\newblock \emph{Russ J Phys Chem+} \textbf{57} (1983) 810.

\bibitem{Pottlacher}
Pottlacher G., J{\"a}ger H.
\newblock \emph{J Non-Cryst Solids} \textbf{205} (1996) 265.

\bibitem{Nikolaev}
Nikolaev D.N., Ternovoi V.Y., Pyalling A.A., Filimonov A.S.
\newblock \emph{Int J Thermophys} \textbf{23} (2002) 1311.

\bibitem{Ternovoi}
Emelyanov A.N., Nikolaev D.N., Ternovoi V.Y.
\newblock \emph{High Temp - High Press} \textbf{37} (2008) 279.

\bibitem{Young}
Young D.A., Alder B.J.
\newblock \emph{Phys Rev A} \textbf{3} (1971) 364.

\bibitem{Bushman}
Bushman A.V., Fortov V.E.
\newblock \emph{Phys-Usp+} \textbf{140} (1983) 177.

\bibitem{Likalter}
Likalter A.A.
\newblock \emph{Phys-Usp+} \textbf{170} (2000) 831.

\bibitem{Martynyuk1998}
Martynyuk M.M.
\newblock \emph{Russ J Phys Chem+} \textbf{72} (1998) 19.

\bibitem{Martynyuk1999}
Martynyuk M.M., Tamanga P.A.
\newblock \emph{High Temp-High Press} \textbf{31} (1999) 561.

\bibitem{Blairs}
Blairs S., H. A.M.
\newblock \emph{J Colloid Interf Sci} \textbf{304} (2006) 549.

\bibitem{Grosse}
Grosse A.V., Kirschenbaum A.D.
\newblock \emph{J Inorg Nucl Chem} \textbf{24} (1962) 739.

\bibitem{Fortov}
Fortov V.E., Dremin A.N., Leont'ev A.A.
\newblock \emph{High Temp+} \textbf{13} (1975) 984.

\bibitem{Vorobiev}
Apfelbaum E.M., Vorob'ev V.S.
\newblock \emph{J Phys Chem B} \textbf{119} (2015) 11825.

\bibitem{Ziman}
Ziman J.M.
\newblock \emph{Principles of the Theory of Solids}.
\newblock London: Cambridge University Press, 1972.

\bibitem{Apfelbaum}
Apfelbaum E.M.
\newblock \emph{Phys Chem Liq} \textbf{48} (2010) 534.

\bibitem{Redmer}
Redmer R., Reinholtz H., Roepke G., Winter R., Noll F., Hensel F.
\newblock \emph{J Phys-Condens Mat} \textbf{4} (1992) 1659.

\bibitem{Levashov}
Knyazev D.V., Levashov P.R.
\newblock \emph{Phys Plasmas} \textbf{21} (2014) 07330.

\bibitem{Khomkin2014}
Khomkin A.L., Shumikhin A.S.
\newblock \emph{J Exp Theor Phys+} \textbf{118} (2014) 72.

\bibitem{Khomkin2015}
Khomkin A.L., Shumikhin A.S.
\newblock \emph{J Exp Theor Phys+} \textbf{121} (2015) 521.

\bibitem{Khomkin2016}
Khomkin A.L., Shumikhin A.S.
\newblock \emph{Contrib Plasm Phys} \textbf{56} (2016) 228.

\bibitem{Smith}
Banerjia A., Smith J.R.
\newblock \emph{Phys Rev B} \textbf{37} (1988) 6632.

\bibitem{Rose}
Rose J.H., Smith J.R., Guinea F., Ferrante J.
\newblock \emph{Phys Rev B} \textbf{29} (1984) 2963.

\bibitem{Clementi}
Clementi E., Roetti C.
\newblock \emph{Atom Data Nucl Data} \textbf{14} (1974) 177.

\bibitem{Bardeen}
Bardeen J.
\newblock \emph{J Chem Phys} \textbf{6} (1938) 367.

\bibitem{Daw}
Daw M.S., Baskes M.I.
\newblock \emph{Phys Rev B} \textbf{29} (1983) 6443.

\bibitem{Rose1983}
Rose J.H., Smith J.R., Ferrante J.
\newblock \emph{Phys Rev B} \textbf{28} (1983) 1835.

\bibitem{Likalter2002}
Likalter A.A.
\newblock \emph{Physica A} \textbf{311} (2002) 137.

\bibitem{Likalter1997}
Likalter A.A.
\newblock \emph{Phys Scripta} \textbf{55} (1997) 114.

\bibitem{Filippov}
Filippov L.P.
\newblock \emph{Metody rascheta i prognozirovaniya svoistv veshchestv (Methods
  for Calculating and Predicting the Properties of Substances)}.
\newblock Moscow: Moscow State University, 1988.

\bibitem{Vorobiev2009}
Apfelbaum E.M., Vorob'ev V.S.
\newblock \emph{Chem Phys Lett} \textbf{467} (2009) 318.

\bibitem{Onufriev}
Onufriev S.V.
\newblock \emph{High Temp+} \textbf{49} (2011) 205.

\bibitem{Morris}
Morris E.
\newblock \emph{AWRE Report}.
\newblock London: UKAEA, 1964.

\bibitem{Hixson}
Hixson R.S., Winkler M.A., Hodgdon M.L.
\newblock \emph{Phys Rev B} \textbf{42} (1990) 6485.

\bibitem{Rakhel}
Korobenko V.N., Rakhel A.D.
\newblock \emph{Phys Rev B} \textbf{85} (2012) 014208.

\bibitem{Apfelbaum2011}
Apfelbaum E.M.
\newblock \emph{J Chem Phys} \textbf{134} (2011) 194506.

\bibitem{Kaschnitz}
Kaschnitz E., Nussbaumer G., Pottlacher G., J{\"a}ger H.
\newblock \emph{Int J Thermophys} \textbf{14} (1993) 251.

\bibitem{Lang}
Lang G.
\newblock \emph{Z Metallkd} \textbf{68} (1977) 213.

\end{thebibliography}

\appendix 

\end{document}